\documentclass[12pt,draftclsnofoot, perreview, onecolumn]{IEEEtran}
\usepackage{balance}
\usepackage{color}
\usepackage{url}
\usepackage{cite}
\usepackage{threeparttable}

\usepackage{amsfonts}
\usepackage{amsmath}
\usepackage{mathrsfs}
\usepackage{multirow}
\usepackage{booktabs}
\usepackage{setspace}
\usepackage{amssymb}
\usepackage{mdwmath}
\usepackage{ifpdf}
\usepackage[dvips]{graphicx}
\graphicspath{{../}}
\DeclareGraphicsExtensions{.pdf,.jpeg,.png,.eps}

\ifCLASSOPTIONcompsoc
  \usepackage[tight,normalsize,sf,SF]{subfigure}
\else
  \usepackage[tight,footnotesize]{subfigure}
\fi

\newcommand{\mathbfit}[1]{\mbox{\boldmath$#1$\unboldmath}}

\newtheorem{algorithm}{\textbf{Algorithm}}

\ifCLASSOPTIONonecolumn

\fi

\ifCLASSOPTIONtwocolumn

\fi

\begin{document}


\title{A New Class of Multiple-rate Codes Based on Block Markov Superposition Transmission}

\author{Chulong~Liang,
        Jingnan~Hu,
        and~Xiao~Ma,~\IEEEmembership{Member,~IEEE,}
        and~Baoming~Bai,~\IEEEmembership{Member,~IEEE}
\thanks{This work was supported by the 973 Program (No.2012CB316100) and the NSF (No.61172082) of China.}
\thanks{Chulong~Liang, Jingnan~Hu and Xiao~Ma are with the Department of Electronics and Communication Engineering, Sun Yat-sen University, Guangzhou 510006, China (e-mail: lchul@mail2.sysu.edu.cn, hujingn@mail2.sysu.edu.cn, maxiao@mail.sysu.edu.cn).}
\thanks{Baoming~Bai is with State Lab. of ISN, Xidian University, Xi'an 710071, Shaanxi, China (e-mail: bmbai@mail.xidian.edu.cn).}}

\maketitle

\begin{abstract}

Hadamard transform~(HT) as over the binary field provides a natural way to implement multiple-rate codes~(referred to as {\em HT-coset codes}), where the code length $N=2^p$ is fixed but the code dimension $K$ can be varied from $1$ to $N-1$ by adjusting the set of frozen bits. The HT-coset codes, including Reed-Muller~(RM) codes and polar codes as typical examples, can share a pair of encoder and decoder with implementation complexity of order $O(N \log N)$. However, to guarantee that all codes with designated rates perform well, HT-coset coding usually requires a sufficiently large code length, which in turn causes difficulties in the determination of which bits are better for being frozen. In this paper, we propose to transmit short HT-coset codes in the so-called block Markov superposition transmission~(BMST) manner.
At the transmitter, signals are spatially coupled via superposition, resulting in long codes. At the receiver, these coupled signals are recovered by a sliding-window iterative soft successive cancellation decoding algorithm.
Most importantly, the performance around or below the bit-error-rate~(BER) of $10^{-5}$ can be predicted by a simple genie-aided lower bound. Both these bounds and simulation results show that the BMST of short HT-coset codes performs well~(within one dB away from the corresponding Shannon limits) in a wide range of code rates.
\end{abstract}

\section{Introduction}
In practical communication systems, it is frequently required to implement several codes with different code rates. This is critical for wireless communication systems to implement the adaptive coded modulation~\cite{tse2005}, which can lower down the outage probability of a single code and hence allows more efficient use of the scarce bandwidth resources~\cite{Alamouti94}. {It is then desirable to implement codes with all different rates of interest with a pair of encoder and decoder~\cite{Sun07}}. Rate-compatible~(RC) codes are a class of multiple-rate codes, which are usually constructed from a mother code by the use of code modifying techniques such as shortening information bits and extending or puncturing parity-check bits~\cite{Hagenauer88, acikel1999, yazdani2004, ha2004rate, zhang2012}, as summarized in~\cite{Casado2009}.
The rate-compatible codes usually have different code lengths for different code rates. For some applications, the code lengths can be varied in a wide range. For example, Raptor codes~(with length as long as required) were optimized by designing the degree profiles to maximize the average throughput~\cite{Uppal11}.
In 2009, Casado~{\em et al} proposed multiple-rate codes with fixed code length by combining rows of the parity-check matrix of a mother code. In 2012, Liu~{\em et al} constructed multiple-rate nonbinary low-density parity-check~(LDPC) codes with fixed block length by using higher order Galois fields for codes of lower rates~\cite{liu2012}. Although fixed-length multiple-rate codes are not applicable to hybrid automatic repeat request~(HARQ) with incremental redundancies, they may find applications in some other scenarios.
{\color{black}
\begin{itemize}
  \item Fixed-length multiple-rate codes can be applied to the scenarios where a framing constraint is imposed on the physical layer. This occurs, for example, when orthogonal frequency division modulation~(OFDM) with a fixed number of subcarriers is used~\cite{Huang08}. In this scenario, fewer information bits should be encoded to maintain the reliability requirement when the channel is in a low signal-to-noise ratio~(SNR) level.
  \item Fixed-length multiple-rate codes may also find applications in flash memory systems, where the channel capacity decreases with ages but the number of memory cells keeps unchanged~\cite{Huang2013}.
      The older flash memory, which has undergone a large number of erase cycles, has a degraded capacity and requires a code of lower rate; while the ``younger" flash memory has a larger capacity and supports a code of higher rate. In this scenario, fixed-length multiple-rate codes with the same pair of encoder and decoder are preferred.
\end{itemize}

Block Markov superposition transmission~(BMST) is a construction of big convolutional codes from short codes~(referred to as \emph{basic codes})~\cite{Ma13,Ma13x}, which has a good performance over the binary-input additive white Gaussian noise channel~(BI-AWGNC).
{\color{black}In~\cite{Ma13x}, it has been pointed out that any short code with fast encoding algorithm and soft-in soft-out~(SISO) decoding algorithm can be chosen as the basic code.}
A sliding-window decoding~(SWD) algorithm with a tunable decoding delay, as similar to the pipeline message-passing decoding algorithm of the LDPC convolutional codes~\cite{Felstrom99,Ueng12}, can be implemented to decode the BMST system.
Most importantly, the performance of the SWD algorithm in the low error rate region can be predicted by a simple genie-aided lower bound.
}

{\color{black}In this paper, the BMST is used to construct a new class of multiple-rate codes with fixed code length.}
First, we propose to construct short multiple-rate codes by adjusting the set of frozen bits in Hadamard transform~(HT), resulting in {\em HT-coset codes}. This family of codes, which have length $N=2^p$ for some $p>0$ and dimension $K$ ranging from $1$ to $N-1$, can be implemented by the use of a pair of encoder and decoder with complexity of order $O(N \log N)$. Such a pair of encoder and decoder were first demonstrated using normal graph for Reed-Muller~(RM) codes by Forney in~\cite{Forney2001}. The HT-coset codes considered in this paper are also closely related to the polar codes~\cite{Arikan2009}, however, channel polarization plays a less important role here because the considered codes will be very short~(typically $N \leq 16$). Then, to improve the performance of the short HT-coset codes, we propose to transmit short HT-coset codes in the BMST manner~\cite{Ma13,Ma13x}.
The resulting system, referred to as a BMST-HT system for convenience, takes as the basic code a Cartesian product of short HT-coset codes and hence has encoding/decoding complexity linearly growing with the transmission block length. {\color{black}In addition, the unified structure of the HT-coset codes can significantly simplify the hardware implementation.} Simulation results show that the BMST-HT systems perform well~(within one dB away from the corresponding Shannon limits) in a wide range of code rates.

{\color{black}The rest of this paper is organized as follows. In Section~\ref{sec:HTcosetCode}, we present the encoding and decoding algorithm for the multiple-rate HT-coset codes. In Section~\ref{sec:BMST-HT}, we construct BMST-HT systems with a general design procedure and present the encoding and decoding algorithm for the multiple-rate BMST-HT systems.
Section~\ref{sec:result} concludes this paper.}

\section{Short Multiple-rate Codes}~\label{sec:HTcosetCode}
\subsection{Fast Hadamard Transform~(FHT)}
\begin{figure}[t]
\centering
\includegraphics[width=0.5\textwidth]{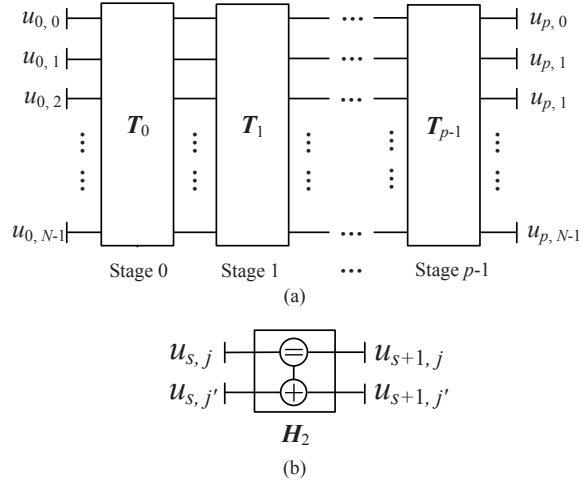}
\caption{Fast Hadamard transform. (a)~A general framework with $N=2^p$. (b)~The basic building block of the Hadamard transform at the $s$-stage, where $(j, j')$ is an $s$-complementary pair, meaning that their binary expansions differ only in the $s$-th bit.}
\label{HadaGraph}
\end{figure}

We consider the Hadamard transform~(HT) defined over the binary field.
Let $N = 2^p$ for some positive integer $p$. Any non-negative integer $j < N$ has a binary expansion $j \triangleq \left(j_{p-1} j_{p-2} \cdots j_1 j_0 \right)_2$ in the sense that $j = \sum_{0 \leq s \leq p-1}j_s 2^s$ and $j_s \in \left\{ 0, 1 \right\}$. We call $\left(j, j' \right) \left( j < j' \right)$ an {\em s-complementary pair} if and only if their binary expansions differ only in the $s$-th bit.
For example, $(0, 1)$ is a 0-complementary pair, $(4, 6)$ is a 1-complementary pair and $(3, 7)$ is a 2-complementary pair.
A Hadamard matrix of order $N$ over the binary field $\mathbb{F}_2 = \left\{ 0, 1 \right\}$ can be defined recursively as~\cite{Forney2001}
\begin{equation}
\mathbfit{H}_N = \left[
    \begin{array}{cc}
    \mathbfit{H}_{N/2} & \mathbfit{H}_{N/2} \\
    \mathbf{0}         & \mathbfit{H}_{N/2}
    \end{array}
    \right]
\end{equation}
for $p > 1$ with $\mathbfit{H}_2 = \left[
\begin{array}{cc}
1 & 1 \\
0 & 1
\end{array}
\right]$. Notice that $\mathbfit{H}_2$ is the transpose of $\mathbfit{F}_2$ in~\cite{Arikan2009}.
Let $\mathbfit{u}_{0} = \left( u_{0,0}, u_{0,1}, \cdots, u_{0,N-1} \right) \in \mathbb{F}_2^N$ be a binary vector.
The Hadamard transform of $\mathbfit{u}_{0}$ is defined as $\mathbfit{u}_{p} = \mathbfit{u}_{0}\mathbfit{H}_{N}$, which can be computed recursively in $p$ stages, see Fig.~\ref{HadaGraph} for reference.
\begin{algorithm}{The Fast Hadamard Transform}\label{alg:FHT}

\begin{itemize}
  \item At stage $s = 0, 1, \cdots, p-1$, compute $\mathbfit{u}_{s+1} \triangleq \mathbfit{u}_{s}\mathbfit{T}_{s}$, where $\mathbfit{T}_{s}$ is a linear transform, whose basic operation is the transform defined by $\mathbfit{H}_{2}$ as
\begin{eqnarray*}
    u_{s+1,j}  &=& u_{s,j}, \\
    u_{s+1,j'} &=& u_{s,j} + u_{s,j'}
\end{eqnarray*}
for each $s$-complementary pair $\left( j, j' \right)$, see Fig.~\ref{HadaGraph}(b) for reference.
\end{itemize}
\end{algorithm}



We use $\mathbfit{U}_{0}$ and $\mathbfit{U}_{p}$ to denote a {\em random} vector and its Hadamard transform, respectively. That is, $\mathbfit{U}_p = \mathbfit{U}_0 \mathbfit{H}_N$, or equivalently\footnote{This can be verified by checking that $\mathbfit{H}_N^2 = \mathbfit{I}_N$, the identity matrix of order $N$.}, $\mathbfit{U}_0 = \mathbfit{U}_p \mathbfit{H}_N$.
In this paper, a message associated with a discrete random variable is represented by its probability mass function. Suppose that, for each $j$, $0 \leq j \leq N-1$, the {\em a priori} messages of $U_{0,j}$ and $U_{p,j}$ are available, which are denoted as $P^{a}_{{U}_{0,j}} \left( u \right), u \in \mathbb{F}_2$ and $P^{a}_{{U}_{p,j}} \left( u \right), u \in \mathbb{F}_2$, respectively. We further assume that all $U_{0,j}$~(resp. $U_{p,j}$) are independent. Then we have
\begin{equation}\label{eq:ProbU0}
P_{ \mathbfit{U}_{0} } \left( \mathbfit{u}_{0} \right) \triangleq \Pr \left\{ \mathbfit{U}_{0} = \mathbfit{u}_{0} \right\} = \prod\limits_{0 \leq j \leq N-1} P^{a}_{{U}_{0,j}} \left( u_{0,j} \right)
\end{equation}
and
\begin{equation}\label{eq:ProbUp}
P_{ \mathbfit{U}_{p} } \left( \mathbfit{u}_{p} \right) \triangleq \Pr \left\{ \mathbfit{U}_{p} = \mathbfit{u}_{p} \right\} = \prod\limits_{0 \leq j \leq N-1} P^{a}_{{U}_{p,j}} \left( u_{p,j} \right)
\end{equation}

{\color{black}For any given component~(say $U_{0,j}$) of interest in $\mathbfit{U}$ and $\mathbfit{V}$, the {\em extrinsic message}~(denoted by, say, $P^{e}_{{U}_{0,j}} \left( u \right), u \in \mathbb{F}_2$)} is defined as a probability mass function that is proportional to the likelihood function, i.e.,
\begin{equation}
P^{e}_{{U}_{0,\!j}} \!\left(\! u \!\right) \!\propto\! \sum\limits_{\!\mathbfit{u}_0:u_{0,\!j}\!=\!u\!} P_{ \mathbfit{U}_{p} } \left(\! \mathbfit{u}_0 \mathbfit{H}_{N} \!\right) \!\prod_{\!0 \!\leq\! k \!\leq N\!-\!1\!, k \!\neq\! j} \!P^{a}_{U_{0, k}}(\! u_{0,\!k} \!), u \in \mathbb{F}_2
\end{equation}
and
\begin{equation}
P^{e}_{{U}_{p,\!j}} \!\left(\! u \!\right) \!\propto\! \sum\limits_{\!\mathbfit{u}_p:u_{p,\!j}\!=\!u\!} P_{ \mathbfit{U}_{0} } \left(\! \mathbfit{u}_p \mathbfit{H}_{N} \!\right) \!\prod_{\!0 \!\leq\! k \!\leq N\!-\!1\!, k \!\neq\! j\!} \!P^{a}_{U_{p,\!k}}(\! u_{p,\!k} \!), u \in \mathbb{F}_2.
\end{equation}
Note that the extrinsic message associated with a random variable is irrelevant to the associated {\em a priori} message.
{\color{black}The complexity to compute all {\em exact} extrinsic messages using the above equations is of order $O(N 2^N)$, which, however, can be reduced if approximate extrinsic messages are tolerable.} This can be attained by performing an iterative message passing/processing algorithm over the normal graph as shown in
Fig.~\ref{HadaGraph}(a).
The normal graph has $p$ super nodes, corresponding to $p$ stages, respectively, where the node $\mathbfit{T}_s$ imposes a constraint that $\mathbfit{U}_{s+1} = \mathbfit{U}_s \mathbfit{T}_s$.
Recalling that the basic building block to implement the transformation $\mathbfit{T}_s$ is the transform defined by $\mathbfit{H}_2$, we have the following algorithm, which falls into the general framework given by Forney~\cite{Forney2001}.

%

\begin{algorithm}{The Forward-backward SISO Algorithm}\label{alg:SISO_FHT_DEC}
\begin{itemize}
  \item {\bf Initialization:}~Assume that $P^{a}_{U_{0,j}}(u),u \in \mathbb{F}_2$ and $P^{a}_{U_{p,j}}(u),u \in \mathbb{F}_2(0 \leq j \leq N-1)$ are available. All intermediate variables are initialized to have a uniform distribution over $\mathbb{F}_2$.

  \item {\bf Iteration:}~Perform iteratively the following steps $J$ times for a preset integer $J>0$.

  {\em Backward recursion:}
   For $s = p-1, p-2, \cdots, 0$, for each $s$-complementary pair $(j,j')$, compute
   \begin{eqnarray*}
   P^{e}_{{U}_{s, j}} \left(\! u \!\right) \!&\propto& \!\sum\limits_{u' \in \mathbb{F}_2 } P^{a}_{ {U}_{s,j'} } \left(\! u' \!\right) P^{a}_{ {U}_{s+1,j} } \left(\! u \!\right) P^{a}_{U_{s+1,j'}} \left(\! u\!+\!u' \!\right), \\
   P^{e}_{ {U}_{s,j'} } \left(\! u' \!\right) \!&\propto&\! \sum\limits_{u \in \mathbb{F}_2 } P^{a}_{ {U}_{s,j} } \left(\! u \!\right) P^{a}_{ {U}_{s+1,j} } \left(\! u \!\right) P^{a}_{ {U}_{s+1,j'} } \left(\! u\!+\!u' \!\right)
  \end{eqnarray*}
   and
   update $P^{a}_{{U}_{s, j}}\left( u \right) = P^{e}_{{U}_{s, j}}\left( u \right)$ and $P^{a}_{{U}_{s, j'}}\left( u' \right) = P^{e}_{{U}_{s, j'}}\left( u' \right)$ for $u, u' \in \mathbb{F}_2$.

  {\em Forward recursion:}
  For $s = 0, 1, \cdots, p-1$, for each $s$-complementary pair $(j,j')$, compute
  \begin{eqnarray*}
   P^{e}_{{U}_{s+1,j}} \left(\! u \!\right) \!&\propto&\! \sum\limits_{u' \in \mathbb{F}_2 } P^{a}_{ {U}_{s+1,j'} } \left(\! u' \!\right) P^{a}_{ {U}_{s,j} } \left(\! u \!\right) P^{a}_{U_{s,j'}} \left(\! u\!+\!u' \!\right), \\
   P^{e}_{ {U}_{s+1,j'} } \left(\! u' \!\right) \!&\propto&\! \sum\limits_{u \in \mathbb{F}_2 } P^{a}_{ {U}_{s+1,j} } \left(\! u \!\right) P^{a}_{ {U}_{s,j} } \left(\! u \!\right) P^{a}_{ {U}_{s,j'} } \left(\! u\!+\!u' \!\right)
  \end{eqnarray*}
  and update $P^{a}_{{U}_{s+1, j}}\left( u \right) = P^{e}_{{U}_{s+1, j}}\left( u \right)$ and $P^{a}_{{U}_{s+1, j'}}\left( u' \right) = P^{e}_{{U}_{s+1, j'}}\left( u' \right)$ for $u, u' \in \mathbb{F}_2$.

\end{itemize}
\end{algorithm}

{\bf Remark.}~It can be seen that both Algorithm~\ref{alg:FHT} and Algorithm~\ref{alg:SISO_FHT_DEC} have complexity of order $O(N\log N)$ since the fast Hadamard transform has $p = \log N$ stages and each stage can be decomposed into $N/2$ independent building blocks as specified by $\mathbfit{H}_2$, each of which is for one complementary pair.

\subsection{Multiple-rate Codes Based on Hadamard Transform}
\begin{figure}[t]
\centering
\includegraphics[width=0.5\textwidth]{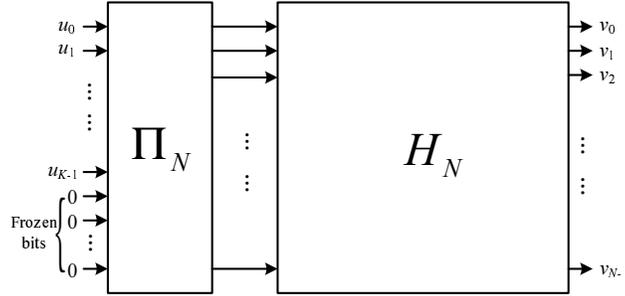}
\caption{The encoding diagram for HT-coset codes.}
\label{Hadamard}
\end{figure}

Given $N$, we can construct a family of codes with dimension $K$ ranging from $1$ to $N-1$. Each code in this family is encoded in the same manner $\mathbfit{v} = \mathbfit{u} \mathbfit{H}_N$. The code rate $K/N$ is attained by fixing some $N-K$ bits in $\mathbfit{u}$ and leaving the other bits free to carry the information. Following~\cite{Arikan2009}, this family of codes are called $\mathbfit{H}_{\!N}$-coset~(or HT-coset) codes. The issue is how to determine a good {\em active} set for a given $K$. For this end, we define a permutation matrix $\mathbfit{\Pi}_N$ of order $N$ such that the rows of the matrix $\mathbfit{G}_N = \mathbfit{\Pi}_N \mathbfit{H}_N$ has a desired order. We may define $\mathbfit{\Pi}_N$ following the polar coding approach~\cite{Arikan2009}, which requires sorting the Bhattacharyya parameters or the mutual information rates for some {\em artificial} channels and usually depends on the channel condition, say, the SNR. Here we define $\mathbfit{\Pi}_N$ according to a channel-independent rule, which was referred to as RM-rule in~\cite{Arikan2009}, such that the rows of $\mathbfit{G}_N = \mathbfit{\Pi}_N \mathbfit{H}_N$ are ordered with non-decreasing Hamming weight. Given $\mathbfit{\Pi}_N$, the encoding algorithm and decoding algorithm for the HT-coset codes
are essentially the same as Algorithm~\ref{alg:FHT} and Algorithm~\ref{alg:SISO_FHT_DEC}, respectively. We only need to freeze some $N-K$ bits when encoding and to initialize these frozen bits with deterministic messages when decoding. These two basic adaptations are summarized in the following algorithms for completeness, where $K$ is specified. See Fig.~\ref{Hadamard} for reference.
\begin{algorithm}{The Encoding Algorithm for HT-coset Codes}
\begin{itemize}
  \item {\bf Input:}~Take as input the information sequence $\mathbfit{u} = (u_0, u_1, \cdots, u_{K-1})$.
  \item {\bf Zero padding and permutation:}~The information sequence $\mathbfit{u}$ is expanded by padding $N-K$ zeros and then permuted by $\mathbfit{\Pi}_N$. The resulting vector is denoted by $\tilde{\mathbfit{u}} = (u_0, u_1, u_{K-1},$ $\underbrace{0, \cdots, 0}_{N-K}) \mathbfit{\Pi}_N$.
  \item {\bf Output:}~Deliver as output the codeword $\mathbfit{v} = \tilde{\mathbfit{u}} \mathbfit{H}_N$, which can be computed by Algorithm~\ref{alg:FHT}.
\end{itemize}
\end{algorithm}
\begin{algorithm}{The SISO Decoding Algorithm for HT-coset Codes}\label{alg:HT-coset-dec}
\begin{itemize}
  \item \textbf{Input:}~Take as input $P_{V_j}^a(v), v \in \mathbb{F}_2$ for $0 \leq j \leq N-1$ and $P_{U_j}^a(u), u \in \mathbb{F}_2$ for $0 \leq j \leq K-1$, where the former are usually computable from the channel observations and the latter are available from the source distribution, which are usually assumed to be uniform distribution.

  \item \textbf{{\em A priori} message freezing:}~Set $P_{U_j}^a(0) = 1$ and $P_{U_j}^a(1) = 0$ for $K \leq j \leq N-1$. 

   \item \textbf{Output:}~Deliver as output $P^{e}_{U_j}(u), u \in \mathbb{F}_2(0 \leq j \leq K-1)$ for hard decisions and $P^{e}_{V_j}(v), v \in \mathbb{F}_2(0 \leq j \leq N-1)$ for possible use in concatenated systems. These extrinsic messages can be computed by Algorithm~\ref{alg:SISO_FHT_DEC} {\color{black}with the exception that the permutation $\mathbfit{\Pi}_N$ need to be taken into account.}
\end{itemize}
\end{algorithm}

\subsection{A Construction Example}\label{example}
\begin{figure}[t]
\centering
\includegraphics[width=0.48\textwidth]{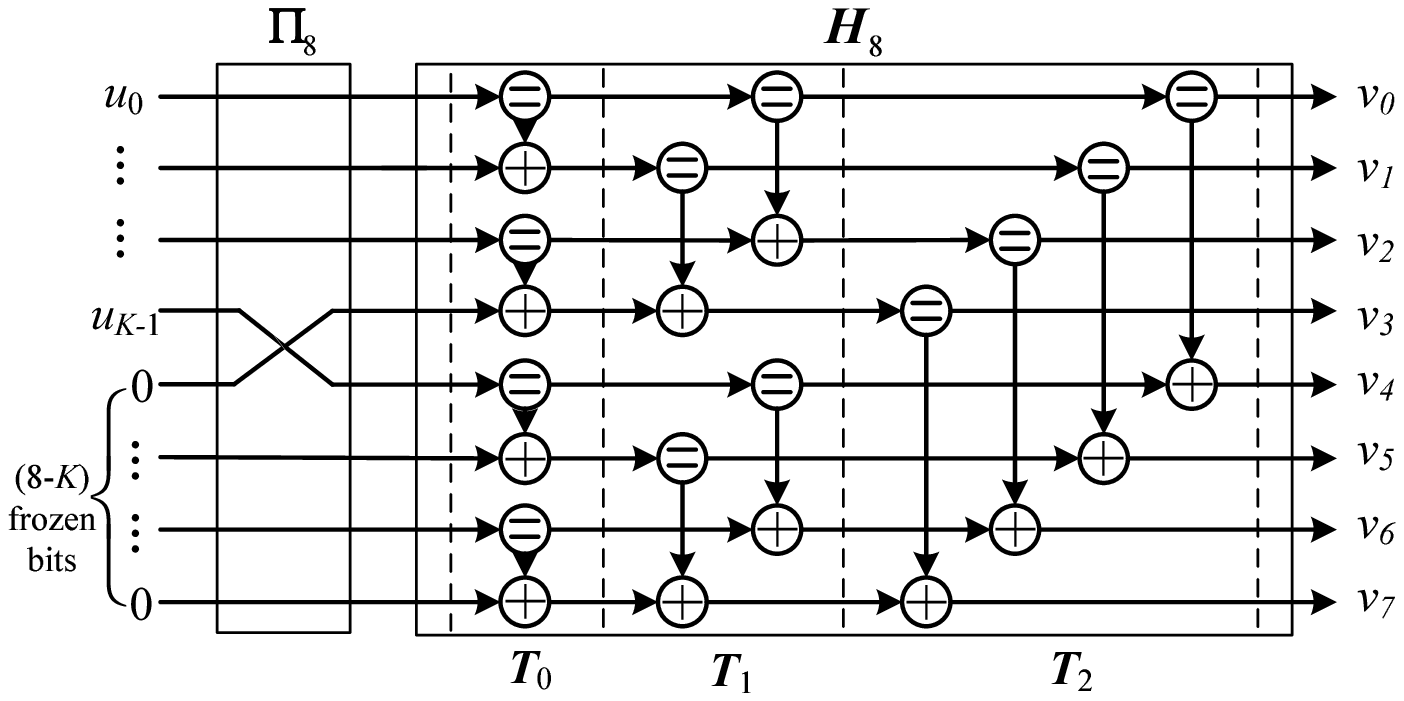}
\caption{The encoding process for HT-coset codes with $N=8$.}
\label{fig:HadaEnc}
\end{figure}

We take $N = 8$ as an example to illustrate the construction procedure. The Hadamard matrix of order $8$ is given by
\begin{equation}\label{HM}
  {\mathbfit{H}_8} = \left[ {\begin{array}{*{20}{c}}
1&1&1&1&1&1&1&1\\
0&1&0&1&0&1&0&1\\
0&0&1&1&0&0&1&1\\
0&0&0&1&0&0&0&1\\
0&0&0&0&1&1&1&1\\
0&0&0&0&0&1&0&1\\
0&0&0&0&0&0&1&1\\
0&0&0&0&0&0&0&1
\end{array}} \right],
\end{equation}
whose rows have Hamming weights $(8, 4, 4, 2, 4, 2, 2, 1)$, respectively. Hence a permutation matrix $\mathbfit{\Pi}_N$ is required to swap the $3$rd row and the $4$th row~(counting from zero). {\color{black}The encoding process is depicted in Fig.~\ref{fig:HadaEnc}, where the dimension $K$ can be varied from $1$ to $7$.}

\begin{figure}[t]
\centering
\includegraphics[width=0.48\textwidth]{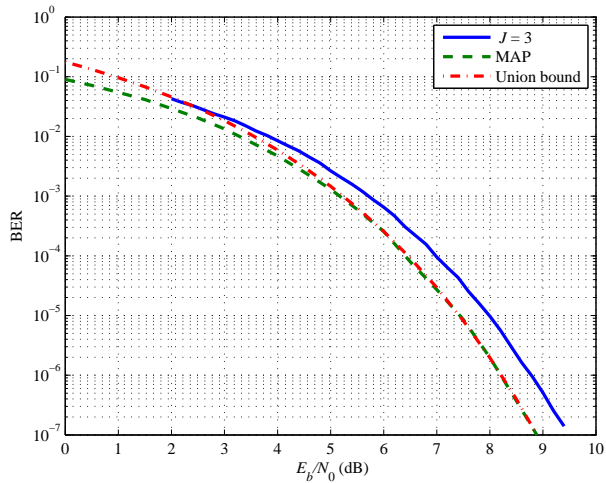}
\caption{The performance of the HT coset code with $K=4$ and $N=8$ under the SISO decoding algorithm~(Algorithms~\ref{alg:SISO_FHT_DEC} and \ref{alg:HT-coset-dec}) with an iteration number $J=3$. The union bound and the performance curve of the MAP decoding are also plotted.}
\label{fig:HTcosetUB}
\end{figure}

{Assume that the codeword is transmitted over an additive white Gaussian noise~(AWGN) channel with binary phase-shift keying~(BPSK) signalling.
{\color{black}The performances of the HT-coset codes at the BER around $10^{-5}$ can be evaluated by simulation, while the performances at the extremely low BER~(say $10^{-10}$) can be predicted using the union bound with the help of the input-output weight enumerating function~(IOWEF)~\cite{Mceliece98,Lin04}.} Since the number of codewords of the considered HT-coset codes with $N=8$ is small, we compute the IOWEFs by an exhaustive search. Table~\ref{tab:HTcosetIOWEF} presents the IOWEFs of all HT-coset codes with $N=8$.
%
We have simulated all these HT-coset codes under the SISO decoding algorithm~(Algorithms~\ref{alg:SISO_FHT_DEC} and~\ref{alg:HT-coset-dec}) with an iteration number $J = 3$. {\color{black}We find that all simulation curves~(not shown here) except that of rate $4/8$ match well with the corresponding union bounds at the BER lower than $10^{-3}$. Fig.~\ref{fig:HTcosetUB} shows the performance curves of the code with rate $4/8$~(the [8, 4] Reed-Muller code). We can see that the performance curve of the rate $4/8$ code under the maximum {\em a posterior} probability~(MAP) decoding matches well with the union bound at the BER lower than $10^{-3}$. However, compared with the union bound, the SISO decoding algorithm~(Algorithms~\ref{alg:SISO_FHT_DEC} and~\ref{alg:HT-coset-dec}) with $J=3$ causes a performance loss of about $0.5$ dB. This performance loss can be narrowed to $0.2$~dB  by increasing $J$, as already pointed out in~\cite{Forney2001}.
}
}
\begin{table}[t]
\renewcommand{\arraystretch}{1.2}
\caption{The IOWEFs for HT-coset Codes with $N=8$\label{tab:HTcosetIOWEF}}
\centering
\begin{tabular}{c||p{7.3cm}}
 \hline
 Codes & IOWEFs \\
 \hline \hline
 $[8, 1]$ & $1 +  XY^8$ \\
 \hline
 $[8, 2]$ & $1 +  XY^4 +  XY^8 +  X^2Y^4$ \\
 \hline
 $[8, 3]$ & $1 + 2XY^4 +  XY^8 + 3X^2Y^4 +  X^3Y^4$\\
 \hline
 $[8, 4]$ & $1 + 3XY^4 +  XY^8 + 6X^2Y^4 + 4X^3Y^4 +  X^4Y^4$ \\
 \hline
 $[8, 5]$ & $1 +  XY^2 + 3XY^4 +  XY^8   + 2X^2Y^2 + 7X^2Y^4 +  X^2Y^6  + 7X^3Y^4 + 3X^3Y^6 + X^4Y^2 + 4X^4Y^4 + X^5Y^4$ \\
 \hline
 $[8, 6]$ & $1 + 2XY^2 + 3XY^4 + XY^8 + 5X^2Y^2 + 8X^2Y^4 + 2X^2X^6 + X^3Y^2 + 12X^3Y^4 + 7X^3Y^6 + 3X^4Y^2 + 11X^4Y^4 + X^4Y^6 + 4X^5Y^4 + 2X^5Y^6 + X^6Y^2$ \\
 \hline
 $[8, 7]$ & $1 + 3XY^2 + 3XY^4 + XY^8 + 9X^2Y^2 + 9X^2Y^4 + 3X^2Y^6 + 3X^3Y^2 + 20X^3Y^4 + 12X^3Y^6 + 9X^4Y^2 + 23X^4Y^4 + 3X^4Y^6 + 12X^5Y^4 + 9X^5Y^6 + 3X^6Y^2 + 3X^6Y^4 + X^6Y^6 + X^7Y^2$ \\
 \hline
\end{tabular}
\end{table}

\section{BMST of Short HT-coset Codes}\label{sec:BMST-HT}
\subsection{Choice of Encoding Memory}\label{subsec:ChooseMemory}
We have demonstrated that Hadamard transform can be used to construct a family of codes with constant length $N$ and flexible dimension $K$.
This family of codes can be implemented by the same pair of encoder and decoder but with adjustable inputs.
However, the performances are far away from the Shannon limits, as evidenced by the construction example with $N=8$. One possible way to improve the performance is to enlarge the order of the Hadamard transform. As $N$ becomes sufficiently large, for any given dimension $K$, we have chance to select active set to approach the channel capacity. The difficulty lies in the choice of the active set when $N$ becomes large. Here we propose to combine the BMST with the HT-coset codes. As pointed out in~\cite{Ma13x}, any short code can be embedded into the BMST system to obtain extra coding gain in the low BER region. The critical parameter for BMST is the encoding memory $m$, which predicts the extra coding gain of $10\log_{10}(m+1)$ dB over AWGN channels.
{\color{black}
Given this predictable extra coding gain, we can use the following deterministic procedure to find the encoding memory $m_K$ required by the BMST-HT system~$(N = 2^p, p > 1)$ to approach the Shannon limit at a target BER $p_{\rm target}$.

{\bf \em A General Procedure of Determining the Encoding Memories for the BMST-HT Systems}\label{pcs:DesignBMST}
\begin{itemize}
  \item For $K = 1, 2, \cdots, N-1$, determine the encoding memory $m_K$ with the following steps.
\begin{enumerate}
  \item From the union bound of the HT-coset code with information bit $K$, find the required $E_b/N_0=\gamma_{K}$ to achieve the target BER $p_{\rm target}$.
  \item Find the Shannon limit for the code rate $K/N$, denoted by $\gamma_{K}^{*}$.
  \item Determine the encoding memory by $10\log_{10}(m+1) \geq \gamma_{K} - \gamma_{K}^{*}$. That is,
        \begin{equation}\label{eq:ComputeMemory}
        m_K = \left\lfloor 10^{\frac{\gamma_K - \gamma^*_K}{10}} - 1 \right\rceil,
        \end{equation}
        where $\lfloor x \rceil$ stands for the integer that is closest to $x$.
\end{enumerate}
\end{itemize}
}

Table~\ref{gap} shows the memory required for each code rate using the BMST of HT-coset codes with $N = 8$ to approach the Shannon limit at a target BER of $p_{\rm target} = 10^{-5}$.

\begin{table}[t]
\renewcommand{\arraystretch}{1.2}
\caption{The Memory Required for Each Code Rate Using the BMST of HT-coset Codes with $N=8$ to Approach the Shannon Limit at the BER of $10^{-5}$\label{gap}}
\centering
\begin{tabular}{p{1.6cm}||c|c|c|c|c|c|c}
 \hline
 \small Rate $R=K/8$ & 1/8 & 2/8 & 3/8 & 4/8 & 5/8 & 6/8 & 7/8\\
 \hline \hline
 \small $\gamma^*_K$~(dB) & -1.2 & -0.8 & -0.3 & 0.2 & 0.8 & 1.6 & 2.9 \\
 \hline
 \small $\gamma_K$~(dB) & 9.6 & 9.8 & 8.4 & 7.7 & 8.9 & 8.6 & 8.2 \\
 \hline
 \small Gap~$\gamma_K\!-\!\gamma^*_K$~(dB)  & 10.8 & 10.6 & 8.7 & 7.5 & 8.1 & 7.0 & 5.3\\
 \hline
 \small Memory $m_K$ & 11 & 10 & 6 & 5 & 5 & 4 & 2 \\
 \hline
\end{tabular}
\end{table}

\subsection{Encoding of BMST-HT System}
\begin{figure}[t]
\centering
\includegraphics[angle=270,width=0.5\textwidth]{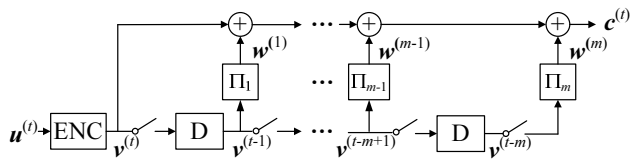}
\caption{The encoding diagram of the BMST-HT system with a maximum memory $m$.}
\label{encoder}
\end{figure}
A BMST-HT system is constructed by taking the encoder for the HT-coset codes as the basic encoder, which accepts as input a binary information sequence of length $BK$ and delivers as output a binary coded sequence of length $BN$. More precisely, the basic code $\mathscr{C}[n,k]$ is a $B$-fold Cartesian product of $\mathscr{C}[N,K]$, where $n=BN$ and $k=BK$. To approach the capacity, we usually choose $B$ such that $BN>10000$. Since $K$ can be varied from $1$ to $N-1$, the basic encoder actually works as a multiple-rate encoder. To guarantee that all members from the family of the considered HT-coset codes have capacity-approaching performance, the encoder must be furnished with  $m~(=\max\limits_{1 \leq K \leq N-1}{m_K})$ interleavers.
The basic structure of the encoder is shown in Fig.~\ref{encoder}, which consists of one basic encoder, represented by ENC, and $m$ interleavers, represented by $\mathbfit{\Pi}_1, \cdots, \mathbfit{\Pi}_m$, respectively. The switches between the buffers D are used to adjust the encoding memory. In our simulations, all interleavers are randomly generated but fixed. The encoding algorithm is described as follows, where $K$ is specified. See Fig.~\ref{encoder} for reference.
\begin{algorithm}{The Encoding Algorithm for BMST-HT System}
\begin{itemize}
  \item \textbf{Input:}~Take as input $L$ blocks of data $\mathbfit{u}^{(0)}, \mathbfit{u}^{(1)}, \cdots, \mathbfit{u}^{(L-1)}$, where $\mathbfit{u}^{(t)} \in \mathbb{F}_2^{BK}$.
  \item \textbf{Memory choosing:}~Turn on the $m_K$ left-most switches and turn off the other $m-m_K$ switches~see Fig.~\ref{encoder} for reference.
  \item \textbf{Output:}~Deliver as output $L+m_K$ coded sub-blocks $\mathbfit{c}^{(0)}, \mathbfit{c}^{(1)}, \cdots, \mathbfit{c}^{(L+m_K-1)}$ for transmission, where $\mathbfit{c}^{(t)} \in \mathbb{F}_2^{BN}$. This can be done by the encoding algorithm of BMST with memory $m_K$~\cite{XiaoMa2013}.
\end{itemize}
\end{algorithm}

\textbf{Remark.} The code rate decreases to $\frac{{KL}}{{N(L + m_K)}}$ due to the termination. However, the rate loss is negligible for large $L$.

\subsection{Decoding of BMST-HT System}
\begin{figure}[t]
\centering
\includegraphics[width=0.5\textwidth]{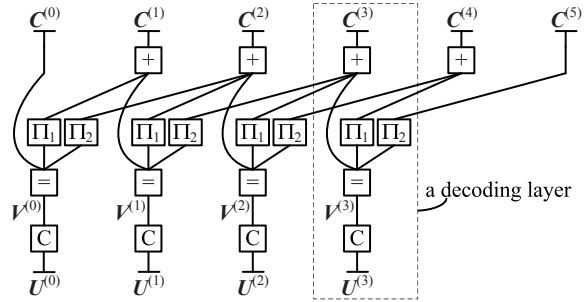}
\caption{The normal graph of the BMST-HT system with $L = 4$ and $m = 2$.}
\label{decoder}
\end{figure}
Assume that $\mathbfit{c}^{(t)}$ is transmitted with BPSK signalling over AWGN channels, resulting in a received vector $\mathbfit{y}^{(t)}$. The decoding algorithm can be implemented as
an iterative message processing/passing algorithm over a unified~(high-level) normal graph, see Fig.~\ref{decoder} for an example of a BMST system with $L = 4$ and $m = 2$. The normal graph can be divided into $L+m$ layers, each of which typically consists of one node of type \fbox{C} for the basic code, one node of type~\fbox{=} that connects to $m$ future layers via interleavers, and one node of type~\fbox{+} that connects to $m$ past layers. The basic structure is applicable to all members from the family of HT-coset codes. Once $K$~(hence the memory $m_K$) is specified, we only need
to disconnect the edges~(if they exist) between the $t$-th layer and the $(t+\ell)$-th layer for $t \geq 0$ and $\ell>m_K$.
{\color{black}With this adaptation, the sliding-window decoding algorithm is the same as Algorithm~3 in~\cite{Ma13x} with $m$ replaced by $m_K$.}

{\bf Remark.}~We need to point out that both the encoding and the decoding have linear complexity in the code length $n$. Actually, since the basic code is a $B$-fold Cartesian product of a short HT-coset code, the computational complexity at the node \fbox{C} is linear in $B$~(equivalently, the basic code length $n$). In addition, the computational complexity at the node~\fbox{$+$} as well as the node~\fbox{$=$} is also linear with the basic code length~$n$.

\subsection{A Construction Example~(Continued)}
We continue the construction example given in Section~\ref{example}. We take the $B$-fold Cartesian product $\mathscr{C}[8,K]^B(1 \leq K \leq 7)$ as the basic code, where $B = 1250$. The memory required for each $K$ to approach the corresponding Shannon limit at the BER of $10^{-5}$ is specified in Table~\ref{gap}. Hence we need an encoder with a maximum memory $m = 11$.
The SISO decoding algorithm~(Algorithms 2 and 4) with an iteration number $J=3$ is used to implement the SISO decoding algorithm for the basic code.
The iterative sliding-window decoding algorithm for the BMST-HT system is performed with a maximum iteration number of $18$, where the entropy-based early stopping criterion~\cite{Ma04,Ma13x} is used with a threshold of $10^{-5}$. Simulation results with $L=1000$ for encoding are shown in Fig.~\ref{polar}, where the decoding delay is specified as $d_K = 2m_K$ for code rate $K/N$. {\color{black}
Also shown in Fig.~\ref{polar} are the genie-aided lower bounds, which are obtained by shifting the corresponding performance curves of the basic codes to left by $10 \log_{10}(1 + m_K)$~dB.}
We can see that the performances of the BMST-HT system match well with the
respective genie-aided lower bounds in the low BER region for all considered code rates.
{\color{black}To evaluate the bandwidth efficiency of the BMST-HT system, we plot the required SNR to achieve the BER of $10^{-5}$ against the code rate in Fig.~\ref{capacity}. We can see that the BMST-HT system achieves the BER of $10^{-5}$ within one dB from the Shannon limit for all considered code rates.}
However, the performance gap of the $[8,4]$ code is slightly larger. This is due to the sub-optimality of Algorithm~\ref{alg:SISO_FHT_DEC}, as mentioned previously. {\color{black}If an optimal~(locally) SISO algorithm~(the MAP decoding) is used for the $[8, 4]$ code, the performance can be improved, as marked in Fig.~\ref{capacity} by the cross $\mathbf{\times}$.

\subsection{Further Discussions}
We have also constructed a BMST-HT system using the $B$-fold Cartesian product $\mathscr{C}[16,K]^B$ $(1 \leq K \leq 15)$ as the basic code, where $B = 625$.
The required encoding memories can be determined following the procedure described in Section~\ref{subsec:ChooseMemory}, which are shown in Table~\ref{tab:memory16}.
The required SNR to achieve the BER of $10^{-5}$ is plotted against the code rate in Fig.~\ref{fig:capacity16}.
{\color{black}We can see that the BMST-HT system also has a BER of $10^{-5}$ within one dB away from the Shannon limits for all considered code rates.
Using HT-coset codes with larger $N$, we can implement codes with finer code rates.
However, this simple approach can only construct codes with rates of form $K/2^p$.
In practice, other code rates~(say $1/3$ and $4/5$ in the standard of Long Term Evolution~(LTE)) are required~\cite{3GPPR11}.
In this case, we can construct the basic code by combining several HT-coset codes with different rates.
Taking rate $1/3$ as an example, we can use the Cartesian product of the HT-coset codes $\left[[4,1]\times[4,1]\times[4,2]\right]^B$ as the basic code.
}
{
\begin{figure}[t]
\centering
\includegraphics[width=0.5\textwidth]{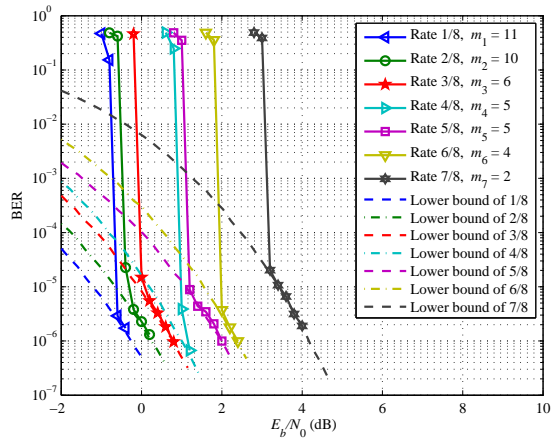}
\caption{\color{black}Performance of the BMST-HT system using the Cartesian products of the HT-coset codes $[8,K]^{1250}(1 \leq K \leq 7)$ with $L=1000$. The sliding-window decoding algorithm is performed with a maximum iteration number $I_{\max}=18$ and a decoding delay $d_K=2m_K$ for the code of rate $K/8$.}
\label{polar}
\end{figure}
}
\begin{figure}[t]
\centering
\includegraphics[width=0.47\textwidth]{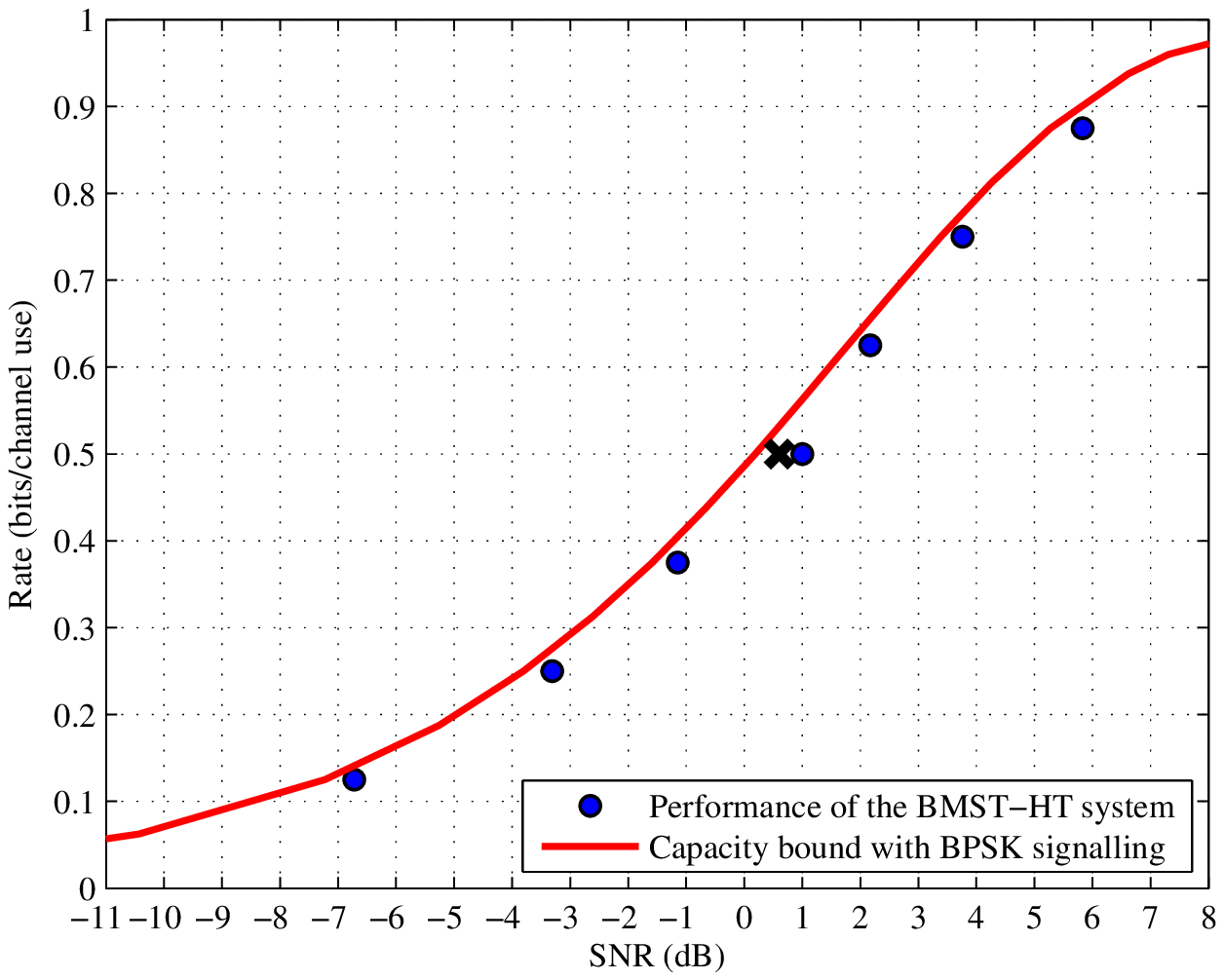}
\caption{The required SNR for the BMST-HT system using the
Cartesian products of HT-coset codes $[8,K]^{1250}(1 \leq K \leq 7)$ to achieve the BER of $10^{-5}$ with BPSK signalling over AWGN channels.}
\label{capacity}
\end{figure}

\begin{table}[t]
\renewcommand{\arraystretch}{1.2}
\caption{The Memory Required for Each Code Rate Using the BMST of HT-coset Codes with $N=16$ to Approach the Shannon Limit at the BER of $10^{-5}$\label{tab:memory16}}
\centering
\begin{tabular}{c||c|c|c|c}
 \hline
 $R=K/16$ & $\gamma^*_K$~(dB) & $\gamma_K$~(dB) & $\gamma_K\!-\!\gamma^*_K$~(dB) & $m_K$ \\
 \hline \hline
 1/16 & -1.4 & 9.6 & 11.0 & 12 \\
 \hline
 2/16 & -1.2 & 9.8 & 11.0 & 12 \\
 \hline
 3/16 & -1.0 & 8.4 & 9.4 & 8 \\
 \hline
 4/16 & -0.8 & 7.7 & 8.5 & 6 \\
 \hline
 5/16 & -0.6 & 7.4 & 8.0 & 5 \\
 \hline
 6/16 & -0.3 & 8.1 & 8.4 & 6 \\
 \hline
 7/16 & -0.1 & 7.9 & 8.0 & 5 \\
 \hline
 8/16 &  0.2 & 7.6 & 7.4 & 4 \\
 \hline
 9/16 &  0.5 & 7.4 & 6.9 & 4 \\
 \hline
 10/16 & 0.8 & 7.2 & 6.4 & 3 \\
 \hline
 11/16 & 1.2 & 7.5 & 6.3 & 3 \\
 \hline
 12/16 & 1.6 & 8.2 & 6.6 & 4 \\
 \hline
 13/16 & 2.2 & 8.4 & 6.2 & 3 \\
 \hline
 14/16 & 2.8 & 8.4 & 5.6 & 3 \\
 \hline
 15/16 & 3.9 & 8.3 & 4.4 & 2 \\
 \hline
\end{tabular}
\end{table}
\begin{figure}[t]
\centering
\includegraphics[width=0.48\textwidth]{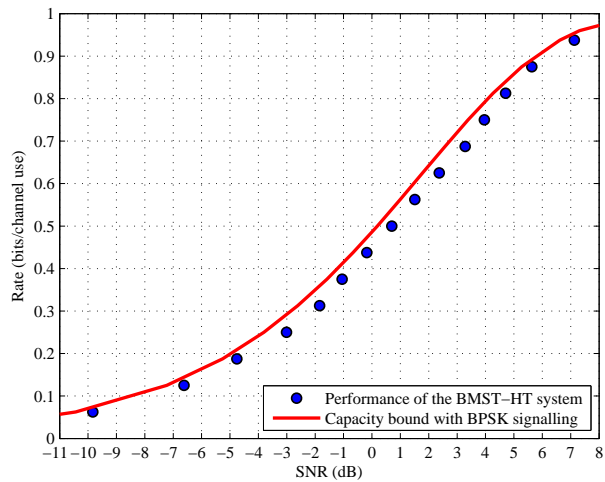}
\caption{The required SNR for the BMST-HT system using the
Cartesian products of HT-coset codes $[16,K]^{625}(1 \leq K \leq 15)$ to achieve the BER of $10^{-5}$ with BPSK signalling over AWGN channels.}
\label{fig:capacity16}
\end{figure}

\section{Conclusions}\label{sec:result}
In this paper, we have proposed a new class of multiple-rate codes by embedding the Hadamard transform~(HT) coset codes into the block Markov superposition transmission~(BMST) system, resulting in the BMST-HT system. The implementation complexity of the system is linear in the code length, while the performance in the low error rate region can be predicted. The simulation results show that the BMST-HT system can approach the Shannon limit within one dB at the BER of $10^{-5}$ for a wide range of code rates.

\balance
\end{document}